\documentclass[aps,prl,twocolumn,floatfix]{revtex4}
\usepackage{graphicx,amsmath}
\usepackage[english]{babel}

\begin{document}

\title{Observation of a Transition from BCS to HTSC-like Superconductivity in Ba$_{1-x}$K$_x$BiO$_3$ Single Crystals}

\author{G.E.~Tsydynzhapov, A.F.~Shevchun, M.R.~Trunin, V.N.~Zverev, D.V.~Shovkun, N.V.~Barkovskiy, L.A.~Klinkova}

\affiliation{Institute of Solid State Physics RAS, 142432
Chernogolovka, Moscow dstr., Russia}

\begin{abstract}
 A study of temperature dependences of the upper critical field
$B_{c2}(T)$ and surface impedance $Z(T)=R(T)+iX(T)$ in
Ba$_{1-x}$K$_x$BiO$_3$ single crystals that have transition
temperatures in the range $6\leq T_c\leq 32$~K (roughly
$0.6>x>0.4$) reveals a transition from BCS to unusual type of
superconductivity. $B_{c2}(T)$ curves corresponding to the
crystals that have $T_c>20$~K have positive curvature (like in
some HTSC), and those of the crystals with $T_c<15$~K fall on the
usual Werthamer-Helfand-Hohenberg curve. $R(T)$ and $X(T)$
dependences of the crystals with $T_c\approx 30$~K and $T_c\approx
11$~K are respectively linear (like in HTSC) and exponential (BCS)
in the temperature range $T\ll T_c$. The experimental results are
discussed in connection with the extended saddle point model by
Abrikosov.

\end{abstract} \maketitle

\section{INTRODUCTION}

Ba$_{1-x}$K$_x$BiO$_3$ compound undergoes a series of phase
transformations on potassium doping. The base composition
BaBiO$_3$ should be a metal with the half-filled conductivity band
according to the band structure calculations~\cite{band}. However,
it turns out to be an insulator due to the formation of a charge
density wave (CDW), which distorts its perovskite lattice to
monoclinic. The insulating state extends up to $x\approx 0.4$,
while the monoclinic symmetry is changed by the orthorhombic one
at $x\approx 0.13$. At the transition into metallic phase
($x>0.4$) the lattice become cubic and no structural transitions
is observed on further increase of $x$~\cite{Pei}. Deformations of
the lattice are small and it remains quasi-cubic for the whole
range of $x$.

Metallic Ba$_{1-x}$K$_x$BiO$_3$ is a superconductor. The
superconducting transition temperature of optimally doped
Ba$_{1-x}$K$_x$BiO$_3$ ($x\approx 0.4$) is rather large ---
$T_c\approx 32$~K, --- and it is usually considered to be a HTSC.
However, Ba$_{1-x}$K$_x$BiO$_3$ contains no transition metal
atoms, has no counterparts for CuO$_2$ planes, and its properties
are isotropic.

One of the striking features that are common for cuprate HTSCs and
Ba$_{1-x}$K$_x$BiO$_3$ is the positive curvature of the upper
critical field temperature dependence $B_{c2}(T)$ as determined by
transport measurements. In the case of optimally doped
Ba$_{0.6}$K$_{0.4}$BiO$_3$ it was observed
in~\cite{Affronte,KBBO}, and, as noted in~\cite{disser}, the
temperature dependences of $B_{c2}(T)$ of
Ba$_{0.6}$K$_{0.4}$BiO$_3$ and
Tl$_2$Ba$_2$CuO$_6$~\cite{Mackenzie} or
Bi$_2$Sr$_2$CaCuO$_8$~\cite{Osofsky}  are coinciding if plotted in
reduced scales. It is well-known that measurements of the upper
critical field of HTSC using different methods produce
inconsistent results; the same is true for
Ba$_{0.6}$K$_{0.4}$BiO$_3$~\cite{Szabo1, Szabo2}.

Another common property of cuprate HTSCs and
Ba$_{1-x}$K$_x$BiO$_3$ is the linear dependence of both components
of the complex surface impedance $Z(T)=R(T)+i\,X(T)$ on
temperature in $T<T_c/2$ range~\cite{Tru1}, which is typical for
$d$-wave symmetry of the order parameter. It is important to bear
in mind that results concerning electromagnetic properties of
Ba$_{1-x}$K$_x$BiO$_3$ are often conflicting because of difficulty
of the production of high-quality samples: even
Ba$_{1-x}$K$_x$BiO$_3$ crystals grown in the same conditions
frequently differ in composition and homogeneity.

In the present paper we study evolution of $B_{c2}(T)$ dependences
for a series of Ba$_{1-x}$K$_x$BiO$_3$ crystals ($0.4<x<0.6$) with
$T_c$ values from 32 to 6~K and find that as $x$ decreases (which
corresponds to increasing $T_c$) a transition from BCS to
HTSC-like superconductivity takes place.  The conclusion is
confirmed by measurements of the surface impedance $R(T)$ and
$X(T)$ in the highest quality crystals of Ba$_{1-x}$K$_x$BiO$_3$
with $T_c=11$~K and $T_c=32$~K.

\begin{figure} \includegraphics[width=\columnwidth]{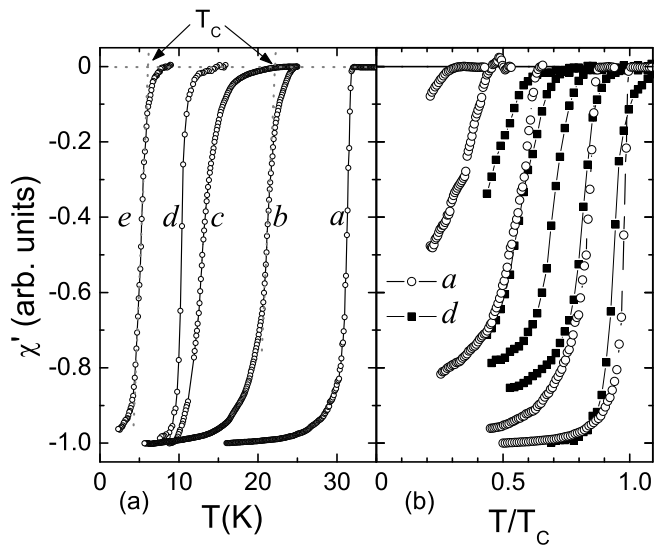}
\caption{ (a) Superconducting transition curves of
Ba$_{1-x}$K$_x$BiO$_3$ crystals obtained by measurements of ac
susceptibility $\chi'(T)$. The definition of the transition point,
which is used in the paper, is shown for samples $b$ and $e$; (b)
the effect of magnetic field $B$ on $\chi'(T)$ curves of the
crystals (circles) $a$ and (squares) $d$. The field values (from
left to the right) for sample $a$ : $B = 0$, 1, 6, 12 and 19.5~T;
for sample $d$: $B=0$, 0.067, 0.133, 0.2 and 0.27~T. Data for
sample $a$ are from the paper~\cite{KBBO}} \label{H0}.
\end{figure}

\begin{figure} \includegraphics[width=0.8\columnwidth]{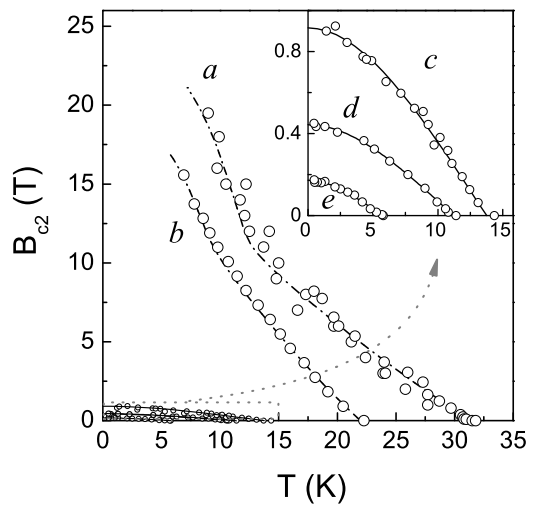}
\caption{ The upper critical field of samples $a-e$. Dash-dot
lines present $B_{c2}(T)$ dependences calculated using the
extended saddle point model ~\protect\cite{Abrikos}. The inset
shows zoom in of the left lower corner of the main graph. Solid
lines are theoretical BCS curves. } \label{Hc2}
\end{figure}

\section{SAMPLES AND EXPERIMENT}

Making of high-quality single crystals of Ba$_{1-x}$K$_x$BiO$_3$
with uniform distribution of potassium at different values of $x$
is a very difficult task. Our samples were grown by the
electrochemical deposition method described
in~\cite{Klinkova1,Klinkova2}. The black with blue-green shine
crystals had roughly cubic shape, well-defined faces, and volume
ranging from 0.2 to 2~mm$^3$. We tried to obtain a series of
crystals with different potassium content covering as wide range
of the superconducting transitions temperatures $T_c$ as possible,
provided that the uniformity of composition within every crystal
is maintained. Since even crystals originating from one batch have
significant variation of $T_c$ (sometimes as large as 10~K), all
samples were preliminary tested by measurements of temperature
dependences of ac susceptibility. Only samples with the narrowest
and most regular superconducting transitions were selected for
further experiments.

Measurement of ac susceptibility $\chi(T)=\chi'+i\chi''$ in
application to superconductors is similar to the investigation of
dc resistivity. In our case the frequency of the ac magnetic field
excitation was 100~kHz and its amplitude was less that 0.1~Oe.
Static external magnetic field up to 17~T was applied using
superconducting magnet perpendicularly to the measuring field.
Samples were not orientated in any special way in respect to the
field, though we checked that results were reproducable for
different positions of the crystals. $T_c$ point was determined
from the transition curve as an intersection of the tangent that
was drawn at the inflexion point and zero level that corresponds
to the normal state (see Fig.~\ref{H0}a).

Temperature dependences of the surface impedance $Z(T)$ were
measured using the resonance technique described in~\cite{Tru2} at
$\omega/2\pi=28.2$~GHz in 0.4-120~K temperature range. $H_{011}$
mode of a cylindrical niobium resonator was employed (the quality
factor of the empty resonator was $\approx 2\cdot 10^6$ in all
operating temperature range); amplitude of the high-frequency
magnetic field was less than 1~Oe. The experimental setup and
method of measurements are described in details in~\cite{Shev}.

 \section{RESULTS}

Temperature dependences of the real part of ac susceptibility
$\chi'(T)$ for all samples with different potassium doping level
that are studied in the present paper are shown in Fig.~\ref{H0}a.
(The transition temperature decreases with increasing $x$.)

The superconductivity is suppressed when magnetic field $B$ is
applied. It is illustrated in Fig.~\ref{H0}b for samples $a$ (the
data for this sample are taken from paper~\cite{KBBO}, that we
published earlier) and $d$. Remarkably, in spite of tremendous,
more than ten times, difference of the field magnitudes $\chi'(T)$
curves of these samples that correspond to the same reduced
temperatures $T/T_c$ have about the same relative transition width
and in general are similar to each other.  This fact is important,
because there is no method to find $T_c$ point from a
non-zero-width transition curve that has a rigorous theoretical
foundation. The method we employ (Fig.~\ref{H0}a) is widespread,
but it is still possible that it introduces a systematic error.
Figure~\ref{H0} strongly suggests that this error, if any, is
identical for different samples.

 \begin{figure} \includegraphics[width=0.75\columnwidth]{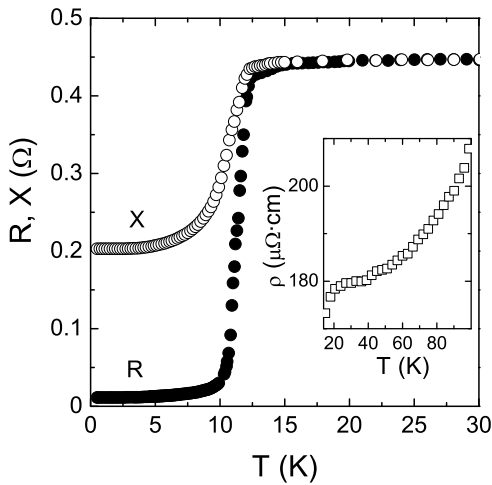} \caption{Surface resistance $R(T)$ and reactance $X(T)$ in superconducting and normal (fragment) states of sample $d$ at 28.2~GHz. The inset shows $\rho(T)$ dependence in $10<T\leq 100$~K range.} \label{RX} \end{figure}

 $B_{c2}(T)$ curves for all samples studied are shown in
Fig.~\ref{Hc2}.  For crystals $c$, $d$, and $e$ with $T_c = 14$,
11, and 6~K  $B_{c2}(0)$ does not exceed 1~T and the curvature of
$B_{c2}(T)$ is negative. Temperature dependences of the upper
critical field of these samples are in complete agreement with the
BCS model. The estimations of the coherence length using formula
$B_{c2}(0)=\Phi_0/(2\pi\xi(0)^2)$ give $\xi(0)=20$, 30, and 40~nm
for samples $c$, $d$, and $e$ respectively.

  $B_{c2}(T)$ curves of crystals $a$ and $b$ ($T_c = 32$ and~22~K)
have positive curvature.  The upper critical fields at $T=0$ are
dozens of times higher than that for crystals $c$, $d$, and $e$,
though available data are insufficient to make even rough
estimation of their magnitude.  At the very least, $B_{c2}(0)$ of
sample $a$ is higher than 25~T, so $\xi(0)$ is no more than 4~nm.
Dash-dot lines in Fig.~\ref{Hc2} are plotted according to the
Abrikosov theory~\cite{abrikos2,Abrikos}. The curves are obtained
by numeric solution of Eq.(15) from~\cite{Abrikos} using the
following parameters (for notation see~\cite{Abrikos}): $\eta =
1$, $\alpha = 4 \pi m_x T_c/\mu_1 m_e = 0.8$ for sample $a$; and
$\eta = 0.9$,
 $\alpha=1.2$ for sample $b$~\cite{note}.

  $B_{c2}(T)$ curves shown in Fig.~\ref{Hc2} demonstrate that the
nature of superconductivity in Ba$_{1-x}$K$_x$BiO$_3$ is changed
with the increase of potassium doping: it turns from an unusual
one to the standard BCS kind.

\begin{figure}
\includegraphics[width=0.8\columnwidth]{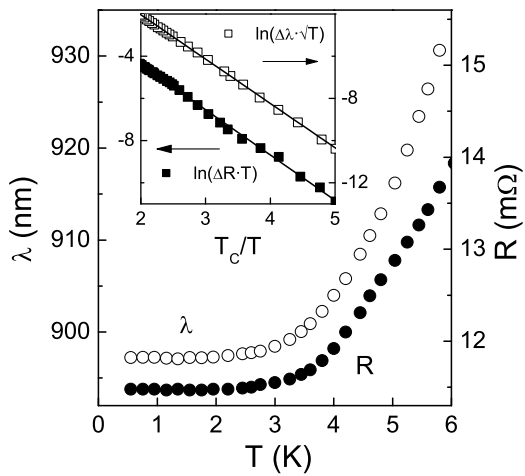}
\caption{ $R(T)$ and $\lambda(T)=X(T)/\omega\mu_0$ of sample $d$
at low temperature. The inset shows logarithms of (solid squares)
$T\Delta R(T)=T[R(T)-R_{res}]$ and (open squares)
$\sqrt{T}\Delta\lambda(T)=\sqrt{T}[\lambda(T)-\lambda(0)]$ in
comparison to (solid lines) the predictions of the BCS model;
their slope gives the gap value $\Delta(0)$.}  \label{lambda}
\end{figure}

Microwave measurements confirm this conclusion. This technique
reveals another important measure of superconducting sample
quality beside the width of superconducting transition --- the
residual surface resistance $R_{res}=R(0)$. It is well-known that
the low-temperature features of the surface impedance $Z(T)$ in
the imperfect crystals are masked by high level of the residual
losses, so they can be observed only in samples with the lowest
values of $R_{res}$.  In the present work, only one of the samples
studied, sample $d$, satisfies this selection criteria.
Temperature dependences of the impedance components for this
crystal are shown in Fig.~\ref{RX}. For $T>T_c$ the normal skin
effect condition $R(T)=X(T)$ is fulfilled, so the temperature
dependence of the resistivity $\rho(T)=2R^2(T)/\omega \mu_0$ can
be derived; it is shown in the inset. Low-temperature parts of
$R(T)$ and $\lambda(T)=X(T)/\omega\mu_0$ curves are shown in
Fig.~\ref{lambda}. In the same way as in classical
superconductors, the resistance $R(T)$ of sample $d$ saturates at
temperature-independent level of the residual losses
$R_{res}\approx 11.5$~m$\Omega$ for $T<T_c/4$ and the field
penetration depth reaches  $\lambda(0)\approx 900$~nm at  $T\to
0$. As shown in the inset of Fig.~\ref{lambda}, both quantities
approach saturation levels exponentially, in complete agreement
with the BCS theory:  $R(T) \propto(1/T)\exp(-\Delta_0/kT)$,
$\lambda(T)\propto(1/\sqrt{T})\exp(-\Delta_0/kT)$, where
$\Delta_0$ --- is the superconducting gap at $T=0$. The values of
$\Delta_0$ derived from these curves agree well and give
$\Delta_0\approx 2.1\,k_BT_c$, which means that the
electron-phonon coupling constant in Ba$_{1-x}$K$_x$BiO$_3$ is not
small. Using well-known BCS equations, we obtain a number of
sample parameters: the relaxation time $\tau\approx 6\cdot
10^{-13}$~s at $T=T_c$, average Fermi velocity $v_F\approx
3\cdot10^5$~m/s (which is about 3 times less than the value
following from the band structure calculations~\cite{band}), and
carrier mean free path $l\approx 180$~nm. According to these
estimations crystal $d$ is a clean London superconductor.

Surface impedance of the optimally doped
Ba$_{0.6}$K$_{0.4}$BiO$_3$ crystal ($T_c\approx 30$~K) was
measured at 9.4~GHz in our paper~\cite{Tru1}. Similar to the case
of tetragonal HTSCs, we observed practically linear temperature
dependences of $\lambda(T)$ and $R(T)$ for
$T<T_c/2$~\cite{Tru1,Tru2}; their extrapolations to $T\to 0$ give
$\lambda(0)\approx 300$~nm and $R_{res}\approx 10$~m$\Omega$.

Hence, microwave measurements confirm that the nature of
superconductivity of samples with $T_c =11$~K and optimally doped
samples is fundamentally different: the former are BCS-type
superconductors, and the latter resemble cuprate HTSCs.

 \section{DISCUSSION}

The origin of the transition from HTSC-like to BCS
superconductivity in Ba$_{1-x}$K$_x$BiO$_3$ compound, which is
revealed in the present paper, should be sought in transformation
of its crystal and electronic structure.

According to the hypothesis~\cite{Klinkova1,Klinkova2},
Ba$_{1-x}$K$_x$BiO$_3$ is formed  from the homologous series of
oxides Ba$_n$Bi$_m$O$_y$ by means of potassium intercalation.
Lattice of the oxides consists of alternating BiO$_2$  and BaO
layers that form a supercell, which size depends on the ratio
$n:m$. Potassium is intercalated into Ba$_n$Bi$_m$O$_y$ between
adjacent BiO$_2$ layers ("a bismuth planes") leading to increase
in the number of Bi ions with the oxidation level $5+$ in these
layers.  As a result Ba$_n$K$_{m-n}$Bi$_m$O$_{3m}$ composition is
formed, meaning that $x$ cannot assume an arbitrary value but
solely those that correspond to a rational fraction $n:m$ of Ba
and Bi content. As shown in~\cite{Amelinckx}, the anisotropic
Ba$_n$Bi$_m$O$_y$ matrix has a mosaic structure at microscopic
level, its lattice consists of ordered blocks that are displaced
by a period of quasi-cubic lattice. When the fraction $n:m$ goes
down, the block size decreases and the layered structure
completely disappears at $n:m=1:2$. If these results are applied
to Ba$_{1-x}$K$_x$BiO$_3$, then we find that a truly isotropic
phase is formed at $x=0.5$ ($n:m=1:2$) instead of $x=0.4$
($n:m=3:5$). Local lattice anisotropy was indeed observed in
Ba$_{0.6}$K$_{0.4}$BiO$_3$ by electron
microscopy~\cite{Klinkova3}. It is not observed in macroscopic
properties probably because of the mosaic lattice structure and
easy twinning, i.e., because of the absence of long-range order in
the positions of potassium.

The arrangement of potassium into layers and related modification
of the oxidation level of Bi ions in the adjacent bismuth layers
leads to modulation of space charge. In the case with the
long-range order (for $n/m \sim 1$) it signifies formation of the
charge density wave (CDW). The very CDW is responsible for
insulating state of Ba$_{1-x}$K$_x$BiO$_3$ for $x<0.4$. As the
size of lattice blocks of Ba$_n$Bi$_m$O$_y$ matrix decreases, the
long-range order disappears, but the "charge-density wave"\ is
preserved at microscopic level defining the short-range ordering
of potassium until the layered structure of the matrix is
completely destroyed. According to this concept, a residual
influence ("traces") of CDW should be retained in the metallic
phase of Ba$_{1-x}$K$_x$BiO$_3$ in the potassium content range
from $x=0.4$ to $0.5$.  Indeed, such traces were observed in the
Compton positron scattering~\cite{Compton} and IR
conductivity~\cite{optics} experiments. Since high-temperature
superconductivity in Ba$_{1-x}$K$_x$BiO$_3$ is observed roughly in
the same composition region, it is possible that the traces of CDW
play a crucial role in its origin.

 A connection between the influence of CDW and superconductivity
may be established within the Abrikosov model~\cite{abrikos2}.
According to it many features of HTSC including the high
transition temperature, $d$-wave pairing symmetry, and positive
curvature of B$_{c2}(T)$ can be explained if the electronic
spectrum contains an extended saddle point (i.e., a flat piece of
the equipotential surface) and the Fermi energy is close to it.
Suitable conditions in Ba$_{1-x}$K$_x$BiO$_3$ can be formed
because of the influence of CDW. According to~\cite{Compton} the
traces of CDW are manifested as a suppressed electron density of
states near the middle of diagonal of the Brillouin zone (point
L); it is exactly where the Fermi surface is located at
$x=0.4$~\cite{band}. Assuming that the suppression forms a flat
part of the spectrum, we can apply the Abrikosov model to explain
all our observations: (i) the linear temperature dependences of
$\lambda(T)$ and $R(T)$ follow from $d$-wave symmetry of the order
parameter, and (ii) the upper critical field curves of samples $a$
and $b$ fit the theoretical dependences well.

The Abrikosov model has a key advantage in respect to application
to Ba$_{1-x}$K$_x$BiO$_3$: the dominating pairing mechanism both
in the Abrikosov and BCS models is the electron-phonon
interaction. Therefore, a transition from one model to another can
occur naturally.

\begin{figure} \includegraphics[width=0.8\columnwidth]{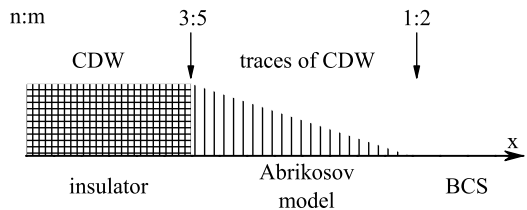}
\caption{The hypothetical phase diagram of Ba$_{1-x}$K$_x$BiO$_3$.
Ba to Bi content ratios $n:m$ that correspond to presumed phase
boundaries are indicated.} \label{ph}
\end{figure}

The framework we propose is illustrated by the diagram in
Fig.~\ref{ph}. At small $x$ the long-range order in positions of
Bi$^{5+}$ ions in the crystals is maintained, so the charge
density wave, which opens a gap at the Fermi surface, is present
and the compound is an insulator.  At the transition into
superconducting metallic phase at $x=0.4$ ($n:m=3:5$) the
long-range order disappears, but, presumably, a short-range
ordering of bismuth ions with different oxidation levels is
preserved. It is manifested in various "traces"\ of the charge
density wave, one of which can be flat parts of the spectrum
formed due to the suppression of the density of states in the L
points in the Brillouin zone. Such flat parts of the spectrum put
into play the Abrikosov model and leads to the high transition
temperature, $d$-wave symmetry of the order parameter (and,
consequently, to linear temperature dependences of $\lambda(T)$
and $R(T)$), and positive curvature of B$_{c2}(T)$. On further
increase of $x$ simultaneously the short-range order is destroyed
and the Fermi surface contracts and move away from the degenerate
point of the Brillouin zone. Because of one or both of these
reasons the Abrikosov mechanism eventually breaks down, but the
electron-phonon coupling make sure that the superconductivity
remains, now within the BCS model. If the latter transition is
caused by complete loss of the short-range order (disappearance of
traces of CDW), then it probably takes place at $n:m=1:2$
($x=0.5$).

\section{CONCLUSIONS}

To conclude, Ba$_{1-x}$K$_x$BiO$_3$ compound is a unique object
exhibiting a transition from HTSC to BCS superconductivity with
the increase of potassium doping $x$ in the metallic phase
($x>0.4$). We observe it clearly in measurements of temperature
dependences of the upper critical field $B_{c2}(T)$ and surface
impedance $Z(T)$ of a series of Ba$_{1-x}$K$_x$BiO$_3$ single
crystals.  The change of the nature of superconductivity can be
due to special features of electron spectrum of
Ba$_{1-x}$K$_x$BiO$_3$ that are related to residual influence of
the charge density wave and employment of the high-temperature
superconductivity mechanism suggested by
Abrikosov~\cite{abrikos2}.

Authors are grateful to V.F.~Gantmakher, A.A.~Golubov, and
V.I.~Nikolaychik for their interest and helpful discussions.

This work is supported by the Russian Foundation for Basic
Research (grants 04-02-17358, 06-02-17098) and scientific programs
of the Russian Academy of Sciences.  G.E.Ts. is grateful to the
Program of Young Scientist Support of the President of Russian
Federation (grant MK-4074.2005.2).).


\begin{thebibliography}{apssamp}
\bibitem{band} S. Sahrakorpi, B. Barbiellini, R.S. Markiewicz
\textit{et al.}, Phys. Rev. B \textbf{61}, 7388 (2000).
\bibitem{Pei} S. Pei, J. D. Jorgensen, B. Dabrowski \textit{et al.}, Phys. Rev. B \textbf{491}, 4126
(1990).
\bibitem{Affronte} M. Affronte, J. Marcus, C. Escribe-Filippine \textit{et al.}, Phys. Rev. B \textbf{49}, 3502 (1994).
\bibitem{KBBO} V.F. Gantmakher, L.A. Klinkova, N.V. Barkovskii \textit{et al.}, Phys. Rev. B \textbf{54}, 6133
(1996).
\bibitem{disser} G.E. Tsydynzhapov, \textit{Ph. D. thesis, Institute of Solid State Physics RAS, 1999}
\bibitem{Mackenzie}A.P.Mackenzie, S.R.Julian, G.G.Lonzarich \textit{et al.}, Phys. Rev. Lett \textbf{71} , 1238 (1993)
\bibitem{Osofsky}M.S.Osofsky, R.J.Soulen,Jr., S.A.Wolf \textit{et al.}, Phys.Rev.Lett. \textbf{71}, 2315 (1993)
\bibitem{Szabo1} P. Szabo, P. Samuely, T. Klein \textit{et al.},
Europhys. Lett. \textbf{41}, 207 (1998).

%P. Szabo, P. Samuely, A.G.M. Jamsem \textit{et al.}, Phys. Rev. B, \textbf{62}, 3502 (2000)

\bibitem{Szabo2} S. Blanchard, T. Klein, J. Marcus \textit{et al.}, Phys. Rev. Lett. \textbf{88}, 177201
(2002).
\bibitem{Klinkova1} L.A.Klinkova, N.V.Barkovskii, S.A.Zver'kov, and D.A.Gusev, Superconductivity {\bf 7}, 1437 (1994)
\bibitem{Klinkova2} L.A.Klinkova, V.I. Nikolaychik, N.V.Barkovskii,\textit{et al.}, Zh. Neorg. Himii \textbf{46}, 715 (2001).
\bibitem{Tru1} M.R.~Trunin, A.A.~Zhukov, G.E.~Tsydynzhapov \textit{et al.}, Pis’ma v ZhETF {\bf 64}, 783
(1996).
\bibitem{Tru2} M.R.~Trunin, Uspekhi Fiz. Nauk \textbf{168}, 931
(1998); \textbf{175}, 1017 (2005).
\bibitem{Shev} A.F.~Shevchun, M.R.~Trunin, to be published in Pribory I Tekhnika Eksper.
(2006).

\bibitem{abrikos2} A.А. Abrikosov, International Journal of Modern Physics \textbf{13},
3405 (1999).

\bibitem{Abrikos} A.А. Abrikosov, Phys. Rev. B \textbf{56}, 5112
(1997).


\bibitem{note} We believe that, because of special features of the model, it is acceptable to use this equation for qualitative comparison with the case of Ba$_{1-x}$K$_x$BiO$_3$, in spite of the fact that the paper~\protect\cite{Abrikos} considers a highly anisotropic material.

\bibitem{Klinkova3} L.A. Klinkova, M. Uchida, Y. Matsui \textit{et al.}, Phys. Rev. B \textbf{67}, R140501
(2003).

\bibitem{Amelinckx} V.I. Nikolaichik, S. Amelinckx, L.A. Klinkova \textit{et al.}, J. SolidState Chem. \textbf{163}, 44 (2002).

\bibitem{Compton} N. Hiraoka, T. Buslaps, V. Homkimaki \textit{et al.}, Phys. Rev. B \textbf{71}, 205106
(2005).

\bibitem{optics} S.H. Blanton, R.T. Collins, K.H. Kelleher \textit{et al.}, Phys. Rev. B \textbf{47}, 996 (1993)


\end{thebibliography}
\end{document}